\documentclass[preprint2]{aastex}

\usepackage{epsfig}
\usepackage{amsmath}
\usepackage{euscript}

\def\thv{\theta_{\text{v}}}

\shorttitle{Perspective on Afterglows}
\shortauthors{Salmonson}

\begin{document}


\title{Perspective on Afterglows: Numerically Computed Views,
Lightcurves and the Analysis of Homogeneous and Structured Jets with
Lateral Expansion}


\author{Jay D. Salmonson}
\affil{Lawrence Livermore National Laboratory, Livermore, CA 94551}



\begin{abstract}
Herein I present numerical calculations of lightcurves of homogeneous
and structured afterglows with various lateral expansion rates as seen
from any vantage point.  Such calculations allow for direct simulation
of observable quantities for complex afterglows with arbitrary energy
distributions and lateral expansion paradigms.  A simple, causal model
is suggested for lateral expansion of the jet as it evolves; namely,
that the lateral expansion kinetic energy derives from the forward
kinetic energy.  As such the homogeneous jet model shows that lateral
expansion is important at all times in the afterglow evolution and
that analytical scaling laws do a poor job at describing the afterglow
decay before and after the break.  In particular, I find that lateral
expansion does not cause a break in the lightcurve as had been
predicted.  A primary purpose of this paper is to study structured
afterglows, which do a good job of reproducing global relationships and
correlations in the data and thus suggest the possibility of a
universal afterglow model.  Simulations of structured jets show a
general trend in which jet breaks become more pronounced with
increasing viewing angle with respect to the jet axis.  In fact, under
certain conditions a bump can occur in the lightcurve at the jet break
time.  I derive scaling relations for this bump and suggest that it
may be a source of some bumps in observed lightcurves such as that of
GRB 000301C.  A couple of lateral expansion models are tested over a
range of efficiencies and viewing angles and it is found that lateral
expansion can, in some cases, substantially sharpen the jet break.  I
show flux surface contour maps and simulated images of the afterglows
which give insight into how they evolve and determine their
lightcurves.
\end{abstract}




\section{Introduction}

It is currently widely believed that gamma-ray bursts (GRBs) derive
from narrow (half-opening angle $\theta_{0,GRB} \sim$ few degrees)
jets of relativistic ejecta pointing toward the observer.  One of the
basic motivations for this has been to relieve the energy crisis in
GRBs by reducing the necessary total energy from the inferred
isotropic equivalent energy (up to several $10^{54}$ ergs) by a factor
$\theta^2_{0,GRB}/2 \sim 10^{-4} - 10^{-6}$ (and thus boosting the
total event rate by the reciprocal, $2/\theta^2_{0,GRB}$, to include
unseen jets directed away from the observer).  If the GRB emission is
collimated, it is plausible that the afterglow shock is also
collimated into a cone with opening angle $\theta_0$, generally thought
to be larger than $\theta_{0,GRB}$.

A rich area of inquiry is then to predict and look for observable
consequences of a narrow jet both in the GRB phase
\cite[e.g.][]{jay00,jay01} and in the afterglow phase
\citep[e.g.][]{rhoads99, sph99}.  The most fundamental consequences
derive from the deceleration of the afterglow as it propagates into
the interstellar medium (ISM).  The relativistic motion of the
emitting shock causes the radiation to be beamed into an angle
$1/\Gamma$ in the observer frame.  At early times, this relativistic
beaming angle is smaller than the physical jet opening angle,
$1/\Gamma < \theta_0$, thus the emission appears to be isotropic.
Eventually, as the jet decelerates, it transitions to $1/\Gamma >
\theta_0$ and the finite, non-isotropic extent of the jet becomes
apparent to both the observer and to the jet itself.  The observer
sees a deficit of flux compared to that expected from an isotropic
emitter.  The jet, now being entirely causally connected, begins to
expand sideways.  These two effects combine to cause a break in the
lightcurve \citep{rhoads99, sph99}.  Roughly twenty such jet-breaks
have been observed \citep[see][and references therein]{fksd01} and
these constitute the most direct evidence for beaming to date.

Also, since the later, slower emission is beamed into a wider angle
than the earlier emission, we expect to see off-axis optical and radio
afterglows where no gamma-ray or X-ray burst was seen
\citep{rhoads97,dgp02}.  To date these so-called ``orphan afterglows''
have yet to be positively observed.

Thus far the study of this afterglow jet-break has assumed an
afterglow pointed directly at the observer with constant, homogeneous,
unstructured energy and mass density $\propto H(\theta_0 - \theta)$
across the jet surface, with a hard edge at $\theta = \theta_0$ where
$H()$ is the Heaviside step function.  As such, from scaling laws,
$t_j \propto \theta_0^{8/3}$, the observation of a jet-break time,
$t_j$, gives direct information about the opening angle of the jet.
While this model is relatively simple to calculate and is amenable to
analytical calculation, it is not necessarily the easiest jet
morphology for nature to produce and thus may not accurately represent
a physical jet.  Firstly, one certainly expects the observers viewing
angle, $\thv$, of the jet to vary.  Furthermore, a homogeneous jet
with a hard edge will be unstable to expansion and rarefaction and is
thus unlikely to propagate intact and is unlikely to have been formed
in the first place.  Recent numerical simulations by \citet{zwm03} of
relativistic jets emerging from stars, within the context of the
`collapsar' model, show substantial structure, with the most energetic
material along the jet axis and decreasing with larger angles from the
axis.

The first semi-analytical calculations of an afterglow jet with
structure, i.e.\ with decreasing energy density and/or Lorentz factor
as a function of angle for the jet axis, was done by \citet{rlr02}
while and analytical treatment was done by \citet{zm02}.  Not long
before, a qualititative discussion of a structured jet model was put
forward by \citet{sg02}.  It was found by \citet{rlr02} and
\citet{zm02} that a universal structured jet, viewed at different
angles, will yield a range of afterglows with jet-break times relating
to viewing angle $t_j \propto \thv^{8/3}$.  By letting the energy per
solid angle of the jet decrease with angle from the jet axis,
$\theta$, as $\epsilon \propto \theta^{-2}$, they were able to
effectively reproduce the observed relation $E \propto t_j^{-1}$
\citep{fksd01}.

In this paper I present calculations that further refine the work by
\citet{rlr02} and \citet{zm02}.  By discretizing the surface of the
afterglow, arbitrary functions of energy
density, Lorentz factor and lateral Lorentz factor (defined in the
fluid frame) can be simulated and lightcurves produced for arbitrary
viewing angles $\theta$.  In so doing, I corroborate some of the key
results of \citep{rlr02,zm02}, i.e.\ $t_j\propto \theta_v^{8/3}$ and
$E_{tot}\propto t_j^{-1}$ if $\epsilon \propto \theta^{-2}$.  Also,
these more detailed calculations allow for a quantitative discussion
of how the lightcurve breaks; in particular I find that a flattening,
and even a bump in the lighcurve is possible just prior to the
jet-break time.  In addition I study lateral expansion and find that
some of the general scaling laws that are widely used in afterglow
work are incorrect.  

\section{Numerical Jet Calculations}

The calculation presented here begins by discretizing the surface of
the afterglow mapped with polar coordinates, $(\theta, \phi)$, into
small elements of solid angle $d\Omega = \cos\psi \sin\theta d\theta
d\phi$ where herein we assume the afterglow is spherical, i.e.~zero
inclination $\psi = 0$, and where $\theta = 0$ corresponds to the jet
axis.  Each surface element plows into the ISM, sweeping up mass
according to its cross-section $dA = R^2 d\Omega$, decelerating,
shocking and radiating.  The physics of this calculation can be broken
up into two parts {\it i)} the dynamical evolution of each surface
element of the afterglow, as dictated by conservation of energy and
momentum and {\it ii)} the radiative mechanism, which I take to be the
standard synchrotron shock model \citep{mr97}.

To calculate the evolution of the afterglow shock one needs only to
invoke conservation of energy and momentum along with an assumption of
radiation losses.  Herein I assume radiative losses are dynamically
insignificant i.e.\ the evolution is adiabatic.  Define the intitial
bulk Lorentz factor $\Gamma_0 = \EuScript{E}_0/\EuScript{M}_0 c^2$ of
a surface element of the afterglow, where $\EuScript{E}_0$ and
$\EuScript{M}_0$ are the initial energy and rest mass per solid angle.
Thus following \citet{pr93} and \citet{rhoads99}, the energy and
radial momentum of a surface element of the afterglow is
\begin{equation}
\begin{split}
\Gamma_0 + f &= (1 + f) \xi \Gamma \\
\Gamma_0 \beta_0 &= (1 + f) \xi \Gamma \beta
\label{E:enmomeqns}
\end{split}
\end{equation}
where $\xi = (E + M)/M$ is the internal energy of the expanding
shell.\footnote{Eqns (\ref{E:enmomeqns}) are the correct equations for
the evolution of jet.  It has been noted by several authors
\citep[e.g.][]{hdl00} that eqns~(\ref{E:enmomeqns}) yield a velocity,
$v \propto R^{-3}$, in the non-relativistic limit that is not
consistent with the Sedov-Taylor (S-T) blastwave solution, $v \propto
R^{-3/2}$, \citep[e.g.][]{shu92b}. Thus it was assumed that these
equations are incorrect for describing afterglow shock dynamics for
the entire evolution.  There have been alternative formulations of the
afterglow shock dynamics.  The reason eqns.~(\ref{E:enmomeqns}) do not
reproduce the S-T solution is because the S-T solution (and the
Blandford-McKee (B-M) solution \citep{bm76}) is a ``blastwave'' and as
such the fireball does work as it expands and thus radial momentum is
not conserved.  The S-T solution depends on the assumption of
spherical symmetry (the lateral expansion of a fluid element is
opposed by expansion of adjacent elements, thus forcing the fluid
radially) and thus does not apply to a jet.  The jet, having nothing
to push on, expands laterally and, thus lacking a radial force, must
conserve radial momentum.  An afterglow jet is more akin to a bullet
fired from a gun than to a spherically expanding blastwave; the jet
and the bullet are isolated, momentum conserving bodies decelerating
due to their interaction with their surroundings, while the blastwave
continues to do work as it expands, by virtue of its sphericity, and
thus is not momentum conserving.  Thus I argue that
eqns.~(\ref{E:enmomeqns}) describe the entire evolution of a jet more
accurately than do the S-T or B-M solutions in their respective
regimes (so long as the jet hasn't expanded so far as to become
spherically symmetric, which I do not find to be the case over the
time interval discussed herein).  Note that the degeneracy of the
energy and momentum equations for $\Gamma \gg 1$ make the behavior of
eqns.~(\ref{E:enmomeqns}) indistinguishable from the B-M solution in
the relativistic regime. }

The mass fraction, $f$, of accumulated interstellar mass density,
$\rho_{ISM} \equiv n m_p$, where $m_p$ is the mass of the proton, is
\begin{equation}
\begin{split}
f &\equiv \frac{\EuScript{M}(R)}{\EuScript{M}_0} = \frac{\Gamma_0 c^2}{\EuScript{E}_0} \int_0^R \rho_{ISM} \frac{\Delta \Omega}{\Delta \Omega_0} r^2 dr
\\ &= \frac{\Gamma_0 m_p c^2}{\EuScript{E}_0 \Delta\Omega_0} \Delta
N_e = 1.9 \times 10^{-54} \frac{\Gamma_0}{\EuScript{E}_{52}
\Delta\Omega_0} \Delta N_e
\label{E:feqn}
\end{split}
\end{equation}
where the number of electrons swept up into the shock element is
\begin{equation}
\Delta N_e = \int_0^R n_0 \Delta\Omega r^2 dr 
\label{E:Nedef}
\end{equation}
and the the energy per solid angle is $\EuScript{E}_{52} =
\EuScript{E}/(10^{52}/4\pi$ ergs$)$, and the element solid
opening angle is
\begin{equation}
\Delta\Omega = \sin\theta \Delta\phi \Delta\theta
\end{equation}
where the position of a surface element, $\theta$, will evolve as the
shock laterally expands due to internal pressure (Section
\ref{sec_sideways_expansion}). The velocity magnitude is $\beta =
|\vec{v}|/c = \sqrt{1 - 1/\Gamma^2}$ and proper time in the fluid
frame is
\begin{equation}
t' = \int_0^R \frac{dr}{\Gamma \beta c} ~~.
\end{equation}
Eqns.~(\ref{E:enmomeqns}) can be solved for the Lorentz factor
\begin{equation}
\Gamma = \frac{\Gamma_0 + f}{\sqrt{1 + 2 \Gamma_0 f + f^2}} ~~.
\label{E:gamma}
\end{equation}
So by specifying $\EuScript{E}_0$ and $\Gamma_0$ and a perscription
for lateral expansion $v_\perp$ (Section
\ref{sec_sideways_expansion}), the entire evolution of the afterglow
as a function of $R$ is determined.

In order to calculate the observed flux, $F_\nu$, at a given
frequency, $\nu$, note that the intensity transforms as $I = I'
\delta^3$ where the Doppler factor is
\begin{equation}
\delta =  [\Gamma (1 - \vec{\beta}\cdot\hat{n})]^{-1} 
\label{E:doppler}
\end{equation}
where $\vec{\beta} = \vec{v}/c$ and $\hat{n}$ is the unit vector
pointing toward the observer.  Following \citet{rlr02}, here I focus
on the power-law branch of the spectrum between the peak, $\nu_m$, and
cooling, $\nu_c$, frequencies.  As such, the proper intensity at the
proper peak frequency, $\nu'_m$, is
\begin{equation}
I'_{\nu'_m} = \frac{P'_{\nu'_m} \Delta N_e}{4\pi R^2 \Delta \Omega}
\label{E:I'def}
\end{equation}
where the shock element has surface area, $R^2 \Delta \Omega$, and the proper
power per electron radiated at, $\nu'_m$, is \citep{wg99}
\begin{eqnarray}
P'_{\nu'_m} = &5.4 \times 10^{-24} \biggl(\frac{\phi_p}{0.59}\biggr) n_0^{1/2} \epsilon_{B,-2}^{1/2} \\ &\times \Gamma \beta^2 ~~ \text{ergs s$^{-1}$ Hz$^{-1}$ electron$^{-1}$} ~.
\end{eqnarray}
where the factor $\beta^2$ has been included here to make this
equation valid in the non-relativistic limit \citep{rl75}.  The
observed intensity at a frequency, $\nu$, is
\begin{equation}
I_{\nu} = I'_{\nu'_m} \biggl(\frac{\nu}{\nu_m}\biggr)^{-\alpha} \delta^3
\label{E:Idef}
\end{equation}
and integrating over the population of radiating electrons, the flux
will go like
\begin{equation}
F_{\nu} = \int \frac{P'_{\nu_m} \Delta A}{4 \pi R^2 \Delta\Omega} \biggl(\frac{\nu}{\nu_m}\biggr)^{-\alpha}  \delta^3  \frac{(1+z)}{D_L^2} \Delta N_e
\label{E:Fluxdef}
\end{equation}
where $D_L$ is the luminosity distance.  The surface area of an
afterglow element as seen by the observer, $\Delta A$, is calculated self
consistently in the code by projecting the elements onto a surface
perpendicular to the observer line of sight.  The minimum electron frequency is
\begin{equation}
\nu'_m = 3.5 \times 10^9 \biggl(\frac{x_p}{0.64}\biggr) \epsilon_{e,-1}^2 n_0 \epsilon_{B,-2}^{1/2} ~ \Gamma^3 ~~\text{Hz}
\end{equation}
and $\nu' = \nu/\delta$ where we take $\nu = 4.4 \times 10^{14}$ Hz
for the R-band.  So 
\begin{equation}
\biggl(\frac{\nu}{\nu_m}\biggr)^{-\alpha} = \biggr[ 7.96 \times 10^{-6} \biggl(\frac{x_p}{0.64}\biggr) \epsilon_{e,-1}^2 n_0 \epsilon_{B,-2}^{1/2} ~ \Gamma^3 \delta \biggr]^{\alpha} 
\end{equation}
and thus eqn.~(\ref{E:Idef}) scales like $I_\nu \propto
\Gamma^{3\alpha+1} \delta^{\alpha+3} R~\nu^{-\alpha}$ $\propto
\Gamma^{3\alpha+2} \delta^{\alpha+4} t~\nu^{\alpha}$, where $t \propto
R/\Gamma/\delta$, thus demonstrating the explicit scalings and
consistency with \citet{rlr02}. For $\alpha = 1/2$ the observed flux
is
\begin{equation}
\begin{split}
F_\nu = &6.4 \times 10^{-57} \biggl(\frac{\phi_p}{0.59}\biggr)
\biggl(\frac{x_p}{0.64}\biggr)^{1/2} \epsilon_{e,-1} n_0
\epsilon_{B,-2}^{3/4}\\  &\times\frac{2}{1+z} \frac{D_L^2(1)}{D_L^2(z)} \int 
\Gamma^{5/2} \delta^{7/2} \beta^2 \frac{\Delta A}{R^2 \Delta\Omega} \Delta N_e ~~\text{mJy}
\label{E:fnu}
\end{split}
\end{equation}
where a cosmology, $(\Omega_b,\Lambda) = (0.3,0.7)$, was used with
$H_0 = 65$ km s$^{-1}$ Mpc$^{-1}$ to give $D_L(1) = 2.2 \times
10^{28}$ cm.  It is important to note that by explicitly evolving
$\Gamma$ (eqn.~\ref{E:gamma}) and $F_\nu$ (eqn.~\ref{E:fnu}) in terms
of the number of swept up electrons, $N_e \propto$ volume
(eqn.~\ref{E:Nedef}), this formulation consistently accomodates
sideways expansion of the jet (Section \ref{sec_sideways_expansion}).
Finally, the flux lightcurve as a function of observer time, $t_{\text{obs}}$, is
calculated by
\begin{equation}
t_{\text{obs}} = (1 + z) \int_0^R (1 - \vec{\beta}\cdot\hat{n}) \frac{dr}{\beta c} 
\label{E:tobs}
\end{equation}
To better compare physical timescales, the redshift dependence is
removed from the times plotted in this paper; $t \equiv
t_{\text{obs}}/(1+z)$.  Thus we have caste the calculation of the
afterglow lightcurve into the state variables of the problem: $R$,
$\Omega$, $\Gamma$, $\delta$.

A calculation proceeds as follows.  An initial afterglow is specified
by $\EuScript{E}(\theta,\phi)$ and $\Gamma_0(\theta,\phi)$ and a
lateral expansion perscription and is allowed to plow into the ISM by
incrementing the radius by $\Delta R/R \sim$ a few percent.  The
intensity, eqn.~(\ref{E:Idef}) and observer time, eqn.~(\ref{E:tobs})
are saved at each surface element.  Thus a lattice of ($I_\nu$,$t$)
pairs are evaluated in ($\theta$,$\phi$,$R$) space.  Intensity is then
interpolated on dataslices of constant observer time $t$.  Finally,
the total flux, $F_\nu$, at each observation is derived from
eqn.~(\ref{E:fnu}) where surface areas, $dA$, are calculated from a
projection of the positions of the observed intensities onto the
observer plane of view.

\section{Review of Homogeneous Jet Model}

\begin{figure}
\plotone{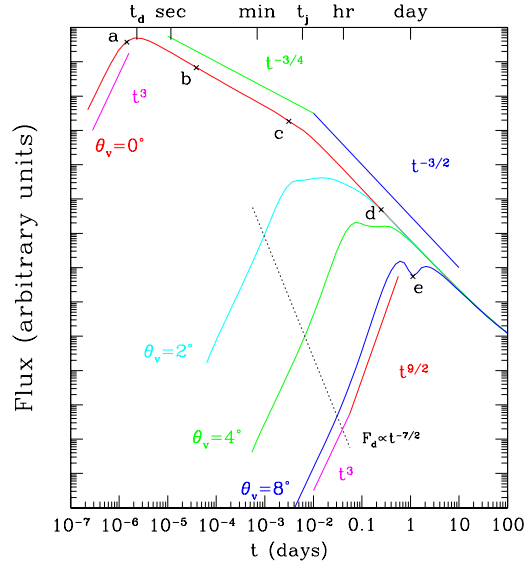} 
\caption{ Shown are calculated lightcurves for several viewing angles,
$\theta_v = 0^\circ,2^\circ,4^\circ,8^\circ$, of an homogeneous
afterglow with opening angle $\theta_0 = 1^\circ$ which plows into an
ISM with number density $n = 1 \text{~cm}^{-3}$.  This afterglow is
given an isotropic equivalent energy of $10^{52}$ ergs and an initial
Lorentz factor $\Gamma_0 = 1000$.  These curves can be compared with
those of \citet{rlr02} and thus flux is in arbitrary units and the big
tick marks correspond to decades.  One can see that the analytical,
power-law solutions for each expansion epoch (Table
\ref{t:simplejet1}) are well reproduced by the numerical calculation.
Also, shown along the top axis, for $\thv=0^\circ$, the theoretically
predicted deceleration time, $t_d \approx 0.2$ s (eqn.~\ref{E:tdec})
and jet-break time $t_j \approx 8.5$ min (Eqn.~\ref{E:tjet})
correspond well to values calculated numerically.  Also shown are
asymptotic relationships for the flux at the deceleration time, $F_{d}
\propto t^{-7/2}$ for $\thv \gg \theta_0$.  It is interesting to note
that extrapolating the $\thv=0^\circ$ curves for the free expansion
epoch, $F \propto t^3$, and the post-jet-break epoch, $F \propto
t^{-3/2}$, as well as the curve $F_{max} \propto t^{-7/2}$, one finds
a common point of intersection.  This physically corresponds to the
lightcurve of a jet of equivalent energy, but with an extremely narrow
opening angle $\theta_0 = 1/\Gamma_0$.
\label{fig:theta1plot}}
\end{figure}

It is worthwhile to begin this discussion with a brief review of the
homogeneous jet model, which is amenable to analytical calculations
and thus allows for comparison and validation of numerical results
with known results.  The apparent surface area of the afterglow goes
like $dA \approx (R~\theta_A)^2$ where $\theta_A$ is the angular size
of the effective viewable aperture onto the afterglow surface
\begin{equation}
\theta_A \approx
\begin{cases}
  1/\Gamma&  \theta_v + 1/\Gamma \ll \theta_0\\ &\quad \text{(relativistic beaming dominated)}\\
  \theta_0&  \theta_v + 1/\Gamma \gg \theta_0\\ &\quad \text{(physical jet extent dominated)}~~.
\end{cases}
\label{E:thetaAdef}
\end{equation}
Also, one can divide the Doppler factor (eqn.~\ref{E:doppler}) into
asymptotic limits
\begin{equation}
\delta \approx 2 
\begin{cases}
  \Gamma&  \theta_v \ll 1/\Gamma + \theta_0 \\
  1/(\Gamma \theta_v^2) & \theta_v \gg 1/\Gamma + \theta_0~~.
\end{cases}
\label{E:dopplerlimits}
\end{equation}
Using eqn.~(\ref{E:fnu}) the flux at a given
frequency is
\begin{equation}
F_\nu \sim \Gamma^{1+3\alpha} \delta^{\alpha+3} R^3 \theta_A^2
\label{E:Fnuscale}
\end{equation}
and from eqn.~(\ref{E:tobs}) the observer time goes like
\begin{equation}
t \approx \frac{R}{\Gamma \delta c} ~.
\label{E:tobsscale}
\end{equation}
Before the afterglow shock has reached its deceleration radius,
$R_{d}$, it coasts freely, $\Gamma \approx \Gamma_0$, and so $\delta
\approx$ const., thus $F_\nu \sim R^3 \sim t^3$, where $t \sim R$.
After the shock passes the deceleration radius, i.e.\ $R > R_{d}$,
the radius and Lorentz factor are related by $R \propto \Gamma^{-2/3}$
(eqns.\ \ref{E:feqn},\ref{E:gamma}).  In this regime there exist
asymptotic power-law slopes for $F(t)$ only in the simple cases where
one of the three angular scales, $\theta_0$,$\theta_v$,$1/\Gamma$,
dominates over the other two.  These three cases are summarized in Table
\ref{t:simplejet1} and can be seen in Fig.~\ref{fig:theta1plot}.  


\begin{table}
\caption{The three asymptotic limits where one of the three angular
scales, $\theta_0$,$\theta_v$,$1/\Gamma$, dominates the other two.
Asymptotic expressions for $\theta_A$ and $\delta$ are given by
eqns.~\ref{E:thetaAdef} and \ref{E:dopplerlimits} respectively.  In
the last column, the expression in parenthesis is for the spectral
slope $\alpha = 1/2$ (eqn.~\ref{E:Idef}). }
\begin{tabular}{cccc}
\tableline 
\tableline \\
\text{limit} & $\theta_A$ & $\delta$ & \text{Flux} ($\alpha=1/2$) \\
\tableline \\
$\theta_0 \gg \theta_v,1/\Gamma$ & $1/\Gamma$ & $2 \Gamma$ & $t^{-3\alpha/2}$ ($t^{-3/4}$) \\
$1/\Gamma \gg \theta_v,\theta_0$  & $\theta_0$ & $2 \Gamma$ & $t^{-3(2\alpha+1)/4}$ ($ t^{-3/2}$) \\
$\theta_v \gg \theta_0,1/\Gamma$  & $\theta_0$ & $2/(\Gamma \theta_v^2)$ & $t^{3(2-\alpha)}$ ($t^{9/2}$) \\
\tableline \\
\end{tabular}
\label{t:simplejet1}
\end{table}


One characteristic timescale of the afterglow model is the
deceleration time
\begin{equation}
t_d = \frac{R_d}{\Gamma \delta c} \approx 0.2 ~ (E_{52}/n)^{1/3} \Gamma_{0,1000}^{-8/3} (1 + \Gamma_0^2 \thv^2) ~\text{s}
\label{E:tdec}
\end{equation}
where I define $R_d = (3 E_0/4\pi\Gamma_0^2 \rho_{ISM} c^2)^{1/3}$
as the radius at which $f = 1/\Gamma_0$ (eqn.~\ref{E:feqn}).  For jets
viewed well off axis, $\thv \gg 1/\Gamma,\theta_0$, the flux
(eqn.~\ref{E:Fnuscale}) varies like $F_{d} \propto
\delta^{7/2}$.  Since $t \propto 1/\delta$, then $F_{d}
\propto t^{-7/2}$.  

Another key timescale is the jet-break time, $t_j$, which occurs when
the shock has decelerated to a Lorentz factor $\Gamma \sim 1/\theta_0$
\begin{equation}
t_j = \frac{5}{4} (\Gamma_0 \theta_0)^{8/3} t_d \approx 8.5~ (E_{52}/n)^{1/3} \theta_{0,1}^{8/3} ~ \text{min} 
\label{E:tjet}
\end{equation}
where $\theta_{0,1} = \theta_0/1^\circ$.  The factor $5/4$ derives
from eqn.~(\ref{E:tobs}) and the fact that the jet-break time occurs
when $1/\Gamma \sim \theta_0$ is observed at the {\em edge} of the jet
rather than at the center.  This is a factor of 5/2 greater than
estimates derived at the jet center \citep[e.g.][]{sph99}.  For $\thv
> \theta_0$ the jet break at time $t_j$ is largely washed out, thus a
more pertinent timescale in this regime is that at which the flux is a
maximum.  This occurs roughly when $\thv \sim 1/\Gamma \gg \theta_0$,
where $\delta \approx 2\Gamma/(1+\thv^2\Gamma^2) \sim \Gamma$ so
$F_{\text{max}} \sim \Gamma^4 \sim t^{-3/2}$, where $t \sim \Gamma^{-8/3}$.

\begin{figure}
\epsscale{1}
\plotone{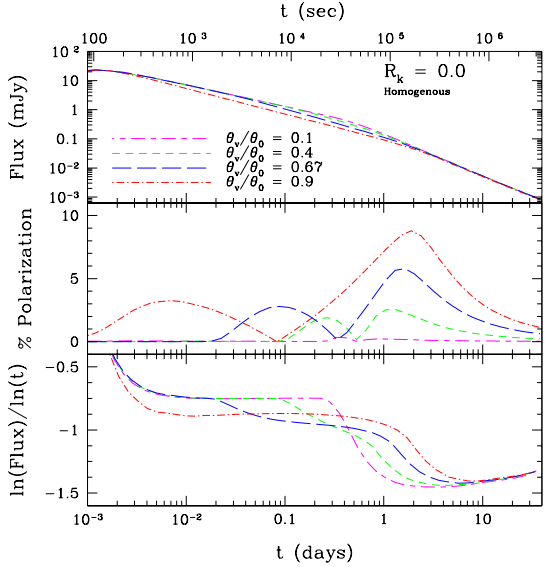} 
\caption{R-band lightcurves (top), polarization curves (middle) and
power-law index (bottom) for a homogeneous jet with opening angle
$\theta_0 = 5^\circ$ and $\EuScript{E} = 10^{52} \text{ergs}/{4\pi}$,
$\Gamma_0 = 100$.  The polarization curves closely match those of
\citet[][Fig. 4]{gl99}.  The first peak of the double-peaked
polarization curves indicates the ``near'' edge of the jet coming into
view, $1/\Gamma > \theta_0 - \thv$, and the second peak corresponds to
the ``far'' edge coming into view, $1/\Gamma > \theta_0 + \thv$.
Therefore comparison of the three figures shows an initial steepening
of the lightcurve for a given viewing angle, $\thv$, corresponding to
the first polarization peak, and additional steepening occurring at
the second peak.
\label{fig:polplot}}
\end{figure}

\begin{figure}
\epsscale{1}
\plotone{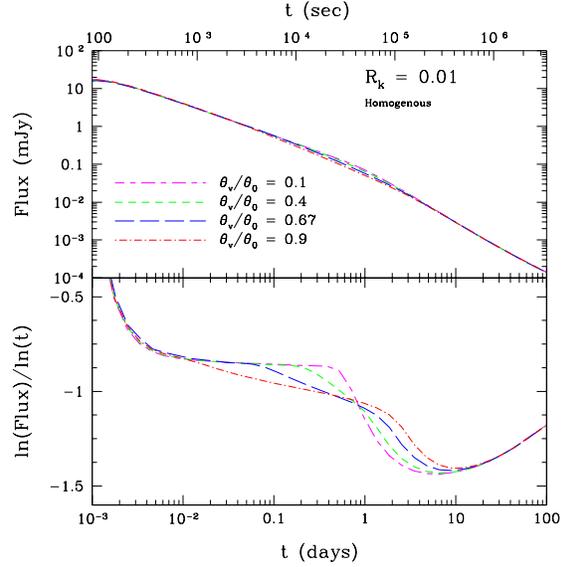} 
\caption{ The same jet as in Fig.~\ref{fig:polplot}, but with sideways
expansion (eqn.~\ref{E:Rkdef}) $R_k = 0.01$, $\gamma'_\perp = 1.6$.
The lightcurves are qualitatively similar to those in
Fig.~\ref{fig:polplot}, with the exception that the lightcurves break
later, at around one day, due to the growth of $\theta_0$ prior to
that time.  The double break is still apparent for larger viewing
angles, $\theta_v$, due to the ``near'' edge of the jet becoming
visible before the ``far'' edge does.}
\label{fig:plplot1.6}
\end{figure}

\begin{figure}
\epsscale{1}
\plotone{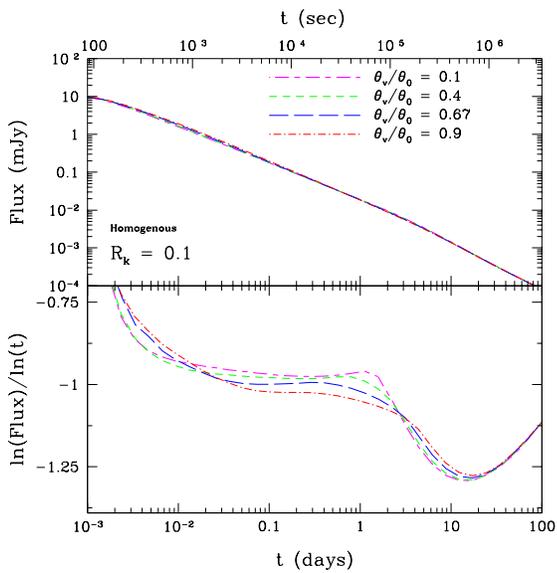} 
\caption{ The same jet as Figs.~\ref{fig:polplot} \&
\ref{fig:plplot1.6}, but with a sideways expansion $R_k = 0.1$,
$\gamma'_\perp = 3.7$.  As in Fig.~\ref{fig:plplot1.6}, the jet break
happens at progressively later times for larger $R_k$ due to the
faster rate of growth of the jet opening angle, $\theta_0$.  The
double break is no longer readily apparent at larger viewing angles,
$\thv$.  Thus the sideways expansion has ``sharpened'' the jet
edge break.  }
\label{fig:plplot3.7}
\end{figure}

\begin{figure}
\epsscale{1}
\plotone{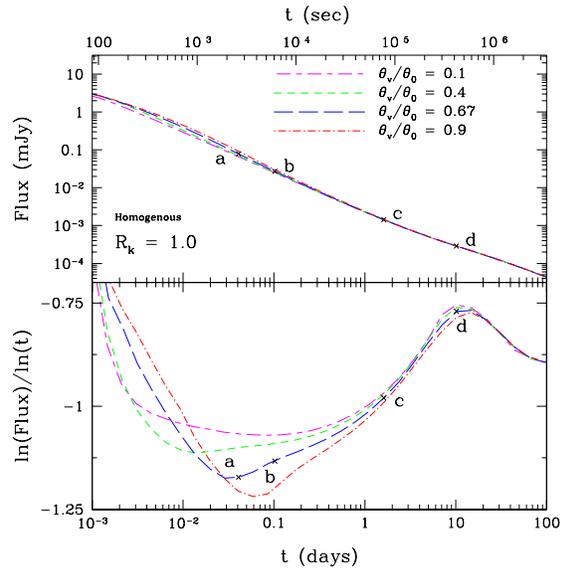} 
\caption{ The same jet as Figs.~\ref{fig:polplot}, \ref{fig:plplot1.6}
\& \ref{fig:plplot3.7}, but with sideways expansion $R_k = 1.0$,
$\gamma'_\perp = 10$.  One can see a signicant pre-jet-break
steepening of the lightcurve circa points a \& b as per Section
\ref{sec_new_laws} and due to Doppler distortions.  At later time,
point d, the lateral expansion is largely exhausted (Section
\ref{sec_sideways_expansion}) and so the decay slope,
$\ln(\text{Flux})/\ln(\text{t}) \approx -3/4$, resembles that of the
case for no lateral expansion (Fig.~\ref{fig:polplot}).  Flux contours
for selected points on curve $\thv/\theta_0 = 0.67$ are shown in
Fig.~\ref{fig:gp10views}. }
\label{fig:plplot10}
\end{figure}

The final phase of the afterglow is when the shock motion becomes
non-relativistic, $\beta \ll 1$.  This regime is beyond the purpose
of this paper, however it is necessary to analytically describe the
non-relativistic lightcurve behavior of the present model so to
understand the asymptotic behavior of the simulations.  For $\beta \ll
1$ eqns.~(\ref{E:enmomeqns}) yield $\beta \propto R^{-3}$,
eqn.~(\ref{E:fnu}) goes like $F_\nu \propto R^3 \beta^2$. and
eqn.~(\ref{E:tobs}) becomes $t \propto R/\beta$. Thus the
non-relativistic lightcurve of the present model is
$F_{\nu,{\text{non-rel.}}} \propto t^{-3/4}$.  Notice that this is
independent of the spectral slope.  This non-relativistic model is
likely incomplete; for instance \citet{dg01} suggest modifying the
shock amplification of the shock-frame magnetic field from $B' \propto
\gamma$ to $B' \propto \sqrt{\gamma (\gamma - 1)}$.  Such
modifications can produce substantially different behavior of the
afterglow lightcurve, including a break at the
relativistic/nonrelativistic transition \citep{hdl00b}.  Since our
understanding of such field generation is still incomplete \cite[for
example][]{rr02}, here I choose not to make any such modifications.
Let it suffice to understand that all post-break lightcurves modeled
in this paper will asympototically approach $F_\nu \propto t^{-3/4}$
as they approach the non-relativistic regime (for example, see late
times of the bottom plot of figs.~\ref{fig:polplot},
\ref{fig:plplot1.6}, \ref{fig:plplot3.7}, \ref{fig:plplot10}).

In order to validate the results presented here, I compared with
existing solutions.  Comparing these homogeneous afterglow solutions
with those of \citet{gps99}, I find very good agreement when one
accounts for three parameter differences.  As mentioned above,
eqn.~(\ref{E:Idef}) is valid for observed frequencies above the
sychrotron frequency, $\nu > \nu_m$.  This simplification is necessary
to remove unwanted spectral breaks when attempting to study the cause
and quality of dynamical jet breaks.  This limit corresponds to the
case when parameter $\phi \gg 1$ in \citet{gps99}.  Secondly,
\citet{gps99} take the velocity of the emitting electrons behind the
shock to be $\Gamma/\sqrt{2}$, where I have assumed it to be $\Gamma$.
I have tested both choices (the former can be implemented by
effectively replacing $\Gamma \rightarrow \Gamma/\sqrt{2}$ in
eqn.~(\ref{E:fnu})) and find a quantitative difference in the flux
surface, but the difference is minimal for the integrated flux; so the
shape of the lightcurves is not significantly affected by this choice,
as can be seen from the results presented in this section.  The last
difference is the choice of spectral index; herein the value $\alpha =
1/2$ is employed while \citet{gps99} use $\alpha = 3/4$.  Adopting the
values used by \citet{gps99}, I find very good agreement with their
results.

\subsection{Lateral Expansion} \label{sec_sideways_expansion}

Thus far the discussion has concerned jets with no lateral (sideways)
expansion, $v_\perp = 0$.  It has been argued that when $c_s t' \gg
\theta_0 c t$, i.e.~when the Lorentz factor has decreased such that
$1/\Gamma > \theta_0$, where $c_s \sim c$ is the sound speed, the
shock will begin expanding sideways \citep{rhoads99,sph99}.  This
increasing angular size of the shock, sweeping up an ever larger
region of ISM, is supposed to impede the forward propagation and the
shock begins to decelerate exponentially with radius.  It is argued
that this shift from forward expansion to lateral expansion will cause
an exponential decay in Lorentz factor as a function of radius,
$\Gamma(R) \propto \exp(-R)$ and thus a break in the afterglow
lightcurve.  Thus one can derive the predicted post-break lightcurve
slope by assuming forward expansion essentially ceases, $R \approx $
constant.  Also, if the jet is narrow, then the jet opening angle will
expand like $\theta \propto t' \propto 1/\Gamma$
\cite[e.g.][]{rhoads99} and since this physical extent of the jet is
growing, one has $\theta_A \sim 1/\Gamma$ (eqn.~\ref{E:thetaAdef}).
Thus from eqn. (\ref{E:Fnuscale}) the flux goes like $F_\nu \sim
\Gamma^{4\alpha +2}$ where $\delta \sim \Gamma$ for $\thv = 0^\circ$.
Also the observer time varies like $t \sim \Gamma^{-2}$
(eqn.~\ref{E:tobsscale}).  Thus $F_\nu \sim t^{-(2\alpha+1)} \sim
t^{-2}$ for $\alpha = 1/2$.


\clearpage
\begin{center}
\epsscale{1.1}
\plotone{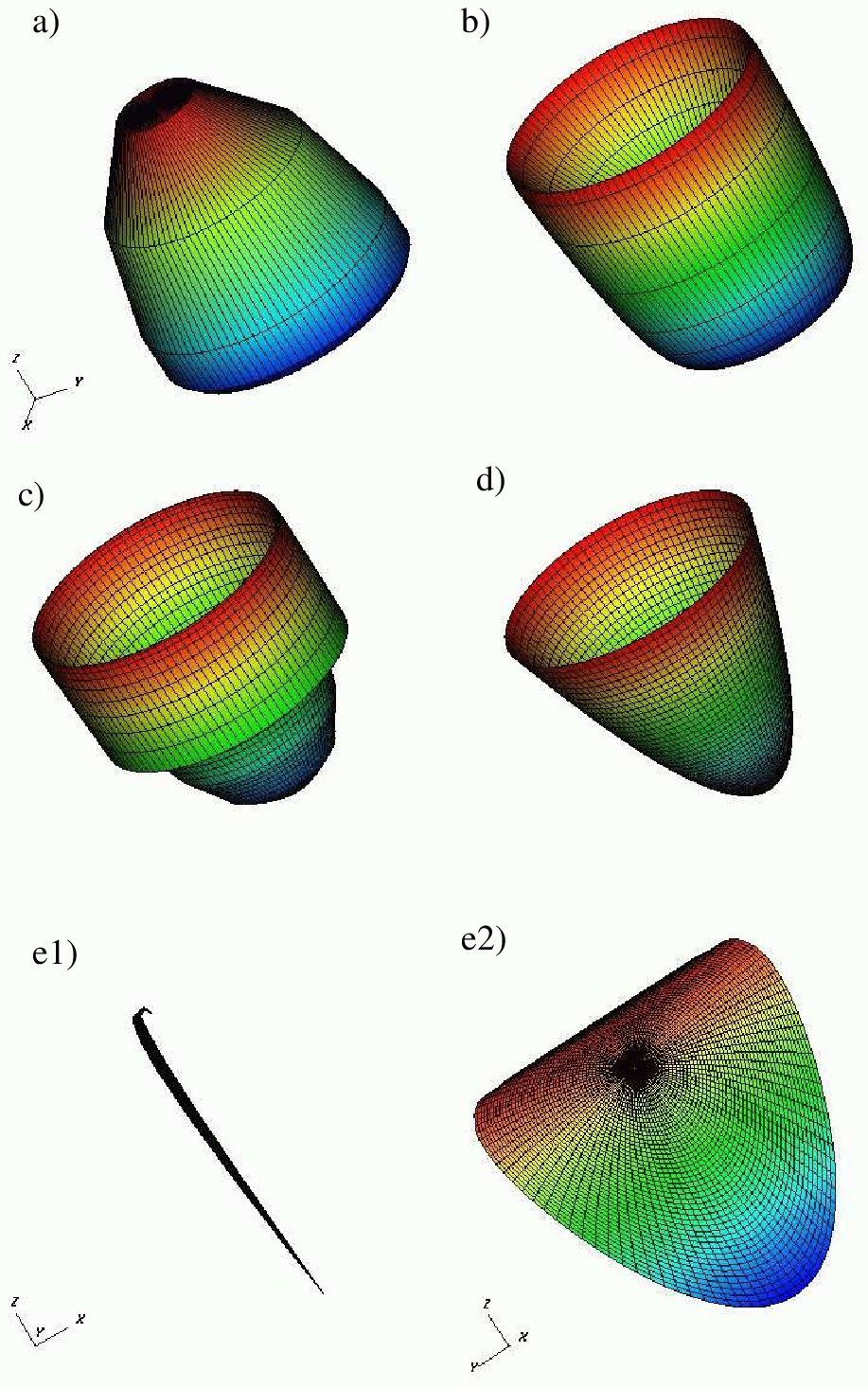} 
\epsscale{1.0}
\begin{figure}
\caption{Views of the flux contour of the afterglow for various points
indicated in Fig.~(\ref{fig:theta1plot}).  The z-axis is the flux
magnitude and the x-y plane is the view plane of the observer, who
would be located in the distance to the upper left.  Flux (z) and
spacial (x and y) axes have been normalized to faciliate viewing.  The
grid lines demarcate the coordinate surface of the afterglow in
$(\theta,\phi)$ at a resolution of $\Delta \theta = 1^\circ/50$,
$\Delta \phi = 360^\circ/100$. At a) the shock, which is poorly
resolved at this resolution, has not reached the deleceration time
(eqn.~\ref{E:tdec}) and thus exhibits a dome-like morphology.  After
passing the deceleration time (b), the flux becomes concave at the
center, as was first described by \citet{gps99}. This concavity is
because the on-axis material is observed to evolve more quickly and
thus decelerate and cool faster than off-axis material. In general,
larger angles from our viewing axis correspond to earlier times in
afterglow evolution and thus higher Lorentz factors, $\Gamma$.
Physically, the rim of maximum flux around the bowl corresponds to the
angle $1/\Gamma$ because material at larger angles having higher
$\Gamma$ and thus tighter beaming, $1/\Gamma$, cannot be seen, and
material at smaller angles corresponding to later times has
decelerated and thus has reduced flux.  The rim thus moves outward as
the afterglow evolves.  Eventually this rim approaches the physical
edge of the jet (c).  As the rim passes the edge of the jet, the light
curve breaks (d).  In (e) we see two views of the afterglow at large
viewing angle, $\thv = 8^\circ$ at a time when most of the flux is
just outside of the rim and thus is basically viewed edge-on by the
observer.  These calculations do not include the thickness of the
shock ($\sim R/\Gamma^2$) and as such one does not expect to observe
such a deep notch in a lightcurve at (e) of
fig.~(\ref{fig:theta1plot}), however, this suggests very high levels
of polarization in orphan afterglows.}
\label{fig:th1views}
\end{figure}
\end{center}
\clearpage

\begin{figure}
\epsscale{1.} 
\plotone{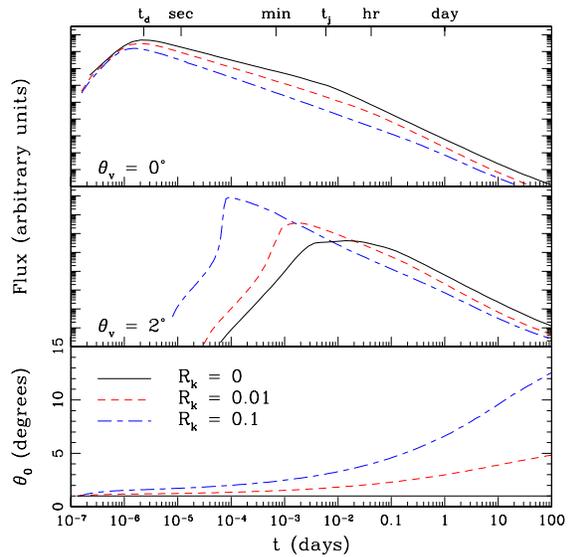} 
\caption{Shown are three different lateral expansion rates for an
afterglow jet of initial opening angle $\theta_0 = 1^\circ$, initial
Lorentz factor $\gamma_0 = 1000$, and isotropic equivalent energy of
$10^{52}$ ergs.  These are $R_k=0$ (no lateral expansion), $R_k=0.01$,
which corresponds to an $\gamma'_\perp = 3.7$, and $R_k = 0.1$, or
$\gamma'_\perp = 10.5$.  The {\it top} figure shows the lightcurves as
seen on axis, $\thv = 0^\circ$, the {\it middle} plot shows
lightcurves viewed at $\thv = 2^\circ$.  Lateral expansion Doppler
boosts the flux into larger angles, $\theta$, from the jet axis and
thus on-axis observers see a reduction in flux (top plot) while
off-axis observers see a surplus of flux (middle plot).  The {\it
bottom} plot show the evolution of the jet opening angle, $\theta_0$,
with observer time.  It is important to notice that {\it none} of
these sideways expansion rates shows marked steepening of the
lightcurve after the jet break time, $t_j$, as has been predicted
analytically (see Sections \ref{sec_sideways_expansion} \&
\ref{sec_new_laws}).  This is because the growth of $\theta_0$ with
time (bottom plot), while rapidly accerating throughout the evolution,
does not become exponential, as analytic arguments would indicate.
\label{fig:th1gpcmpplot}}
\end{figure}

\begin{figure}
\epsscale{1}
\plotone{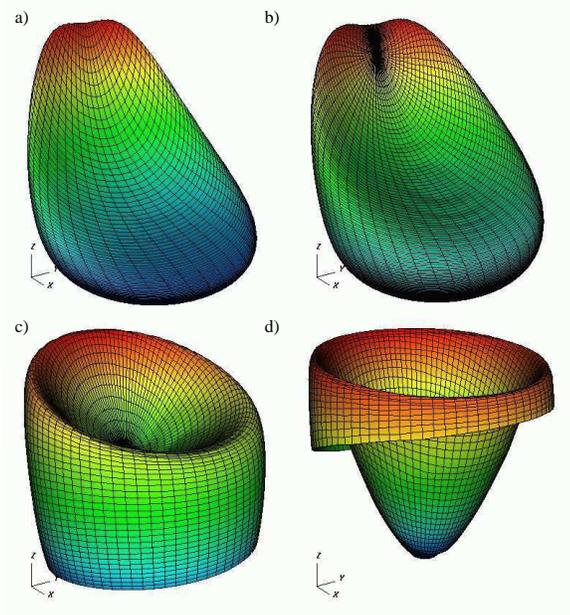} 
\caption{ Flux contours for selected points of Fig.~\ref{fig:plplot10}
for $\thv/\theta_0 = 0.67$. See Fig.~\ref{fig:th1views} for a
description of axes.  Plots a) and b) show flux rim
(e.g.~Fig.~\ref{fig:th1views}b) to be heavily skewed and distorted due
to the Doppler factor of the sideways expansion; material near the
core of the jet is seen to be brighter.  In b) the rim expands beyond
the pole of the jet.  Plot c) shows the rim becoming less distorted as
the jet decelerates, but still sloped, and the edge of the jet is
visible.  Plot d) shows the rim to be very flat and symmetrical as the
jet has decelerated substantially thus sideways expansion and
resultant aberration has subsided.  Grid resolutions are $\Delta
\theta = 5^\circ/150$ and $5^\circ/50$ for a) \& b) and c) \& d)
respectively, with $\Delta \phi = 360^\circ/90$. }
\label{fig:gp10views}
\end{figure}

\begin{figure}
\plotone{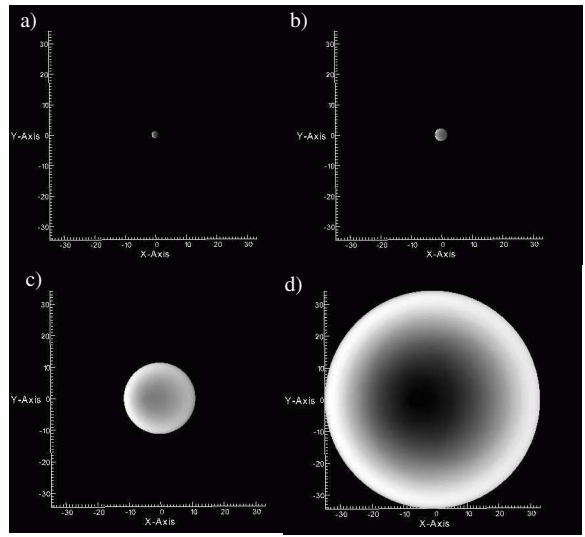} 
\caption{What an afterglow might look like.  Views of the flux
surfaces of the previous Figure \ref{fig:gp10views} along the
z-axis. The doppler brightened core is on the left side of the flux
surface.  Flux magnitude is measured in a normalized color scale;
black is zero, white is the maximum. The x and y axes are shown in
their true scale and measured in lightdays.  By comparing with
Fig.~\ref{fig:plplot10}, one can measure the superluminal expansion of
the jet. }
\label{fig:gp10zviews}
\end{figure}

\begin{figure}
\plotone{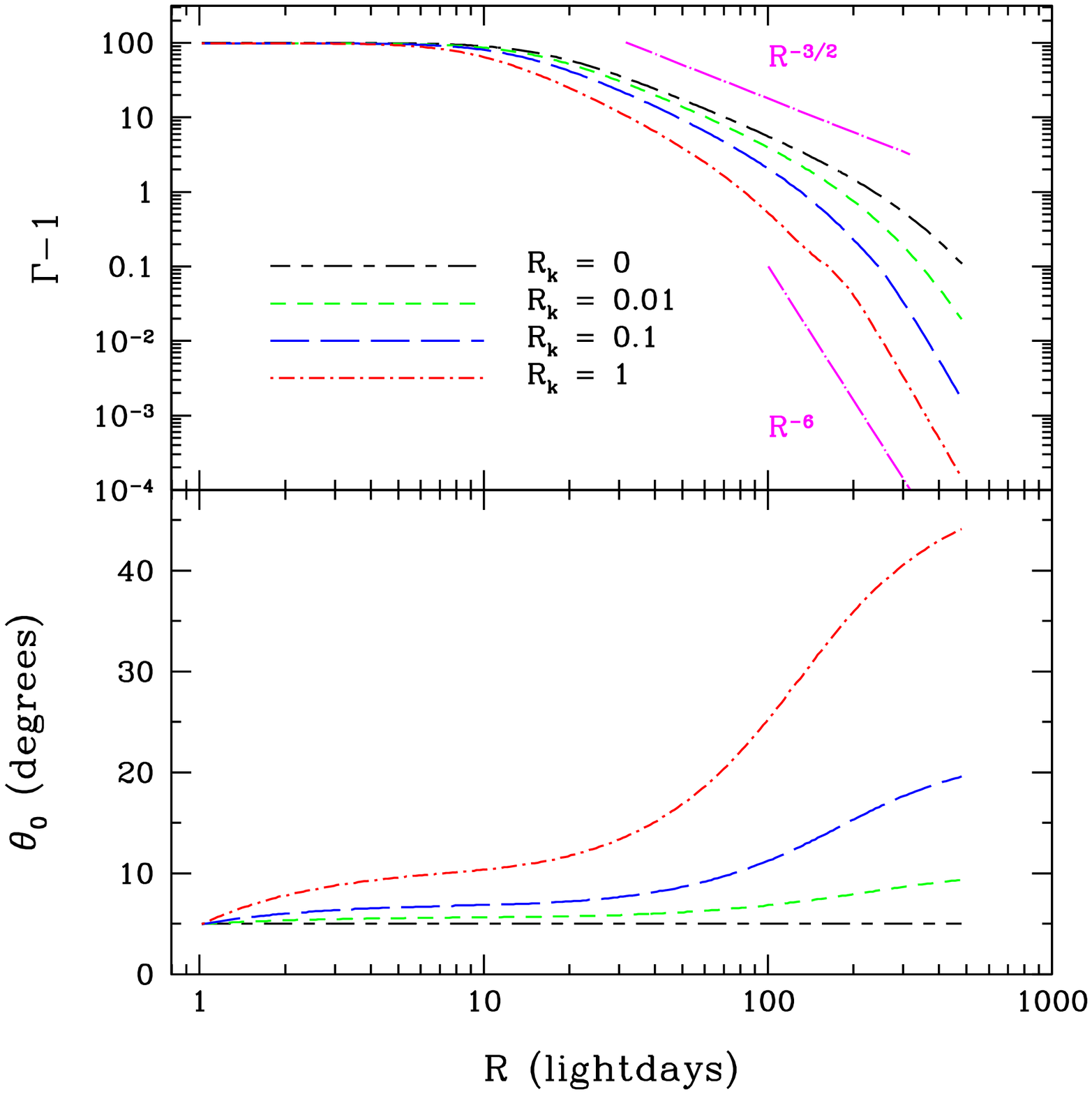} 
\caption{For the homogeneous jet model, the evolution of the jet
specific kinetic energy, $\Gamma-1$, and opening angle, $\theta_0$,
are shown for four jet expansion parameters, $R_k$,
(eqn.~\ref{E:Rkdef}). The kinetic energy (top) does decrease more
rapidly with radius, $R$, for large lateral expansion parameters, but
does not become exponential in $R$ as argued by
\citet{rhoads99}. Instead it asymptotically steepens to the
nonrelativistic limit, $\Gamma-1 \propto R^{-6}$.  For this reason the
sequence of Figs.~(\ref{fig:polplot}, \ref{fig:plplot1.6},
\ref{fig:plplot3.7}, \ref{fig:plplot10}) do not exhibit a
progressively more pronounced break due to lateral broadening of the
jet.  The opening angle (bottom) is seen to expand rapidly once the
jet decelerates to roughly $\theta_0 \sim 1/\Gamma$ as expected, but
because lateral expansion depends on forward expansion, it slows down
to a maximum final opening angle $\theta_{\text{max}} = \theta_0
+\arctan{(\sqrt{R_k})}$.  Since these simulations commence at $R = 1$
lightday (prior to deceleration; $\Gamma \approx $ constant), one can
see some early spreading, $\theta_0 \sim \int (v_\perp/R) dt \sim
\log(R)$.  While this particular behavior is an artifact of starting
the simulation at $R = 1$ lightday, it does indicate that strong
lateral spreading can significantly rearrange the initial jet
morphology before it begins to decelerate, thus hinting at the
possibility of a universal jet shape independent of initial
conditions.}
\label{fig:th0Ngamma}
\end{figure}


However, the assumption that the radial expansion in $R$ effectively
halts as a result of the lateral expansion of $\theta_0$ is an extreme
limit characterized by a narrow opening angle, $\theta_0 \ll
1/\gamma$, and a lateral expansion velocity, assumed constant, that
eventually exceeds the decelerating forward velocity.  Several workers
\citep{wl00,msb00,pm99} have shown that in practice the lateral
expansion of the jet does not reproduce the scaling discussed above
and cannot create a prominent, ``sharp'' break and instead tends to
smooth and broaden the jet-break of eqn.~(\ref{E:tjet}).  Certainly
the approximation of a narrow opening angle does not hold in general
and we presently consider the behavior of $v_\perp$.

The lateral expansion is a result of internal pressure in the shocked
medium which, in turn, is the result of the forward motion of the
shock into the ISM. Thus I argue that a physically motivated model for
lateral expansion is to make the rate of lateral expansion a function
of the rate of forward expansion.  Thus, one does not expect the
lateral expansion velocity to exceed the forward motion of the shock,
from which it derives its energy.  There are two basic reasons for
this.  First, as the shock progresses into the ISM, in decelerates,
thus the rate of injection of internal energy into the shock from bulk
kinetic energy is monotonically decreasing ($\sim \Gamma$).
Furthermore the shock is sweeping up interstellar mass at an
increasing rate ($\sim R^2$).  As was pointed out by \citet{msb00},
this mass has no inherent lateral velocity component and thus new mass
must be constantly accelerated.  This is a significant source
of ``drag'' on the decreasing internal pressure responsible for the
expansion.  Thus the assumption that lateral expansion is at all times
constant is unphysical.

In this section calculations are presented for afterglows with a
simple lateral expansion model and it is found that lateral expansion
serves primarily to smooth the break if it has any effect at all and,
perhaps more importantly, provides very little if any steepening of
the post-break light curve.

Each shock element originally moves radially with a Lorentz factor
$\gamma$.  The relativistic jump conditions imply the shocked gas is
heated by $\sim \gamma - 1$ and is thus driven to expand laterally via
its internal pressure.  The actual lateral expansion that results is a
complex balance between internal pressure forces of the shocked gas
and the ``drag'' of constantly sweeping up and laterally accelerating
fresh ISM.  This balance depends intimately on the hydromagnetic
nature of collisionless shocks and is not well understood as of yet.
Here I simply perscribe that the lateral kinetic energy of the shock,
in its radially comoving frame, is a constant proportion of the radial
kinetic energy.  Thus I define this ratio
\begin{equation}
R_k = \frac{\gamma'_\perp - 1}{\gamma - 1}
\label{E:Rkdef}
\end{equation}
which is a parameter measuring the shocks efficiency at converting
radial kinetic energy, $\gamma - 1$, into lateral kinetic energy,
$\gamma'_\perp - 1$, and thus can vary from zero to unity.

Let us define lateral velocity as the velocity of the shock in the
radially co-moving frame.  This paper assumes axisymmetry of the
afterglow so the proper coordinate velocity, 4-velocity and Lorentz
factor in this frame are denoted respectively; $v'_\perp(\theta)$,
$U'_\perp(\theta) = v'_\perp(\theta) \gamma'_\perp(\theta)$,
$\gamma'_\perp = (1 - {v'}^2_\perp (\theta))^{-1/2}$.  Let us assume
the shock expands uniformly, i.e.~$v_\perp(\theta) \propto \theta$
(see Sec.~\ref{section:nonuniform} for a relaxation of this
condition), thus noting that $R_k$ goes like energy, given a ratio,
$R_{k0}$, at the edge of the jet, $\theta = \theta_0$, this ratio at
all angles, $\theta$, will be determined by
\begin{equation}
R_k \equiv \biggl(\frac{\theta}{\theta_0}\biggr)^2 R_{k0} ~.
\label{E:Rkeqn}
\end{equation}
It is a convenient fact that since the radial velocity, $v$, and
lateral velocity, $v'_\perp$, are orthogonal in the shock frame, the
Lorentz factor of the shock in the lab frame is
\begin{equation}
\Gamma = \gamma \gamma'_\perp ~.
\label{E:Gammadef}
\end{equation}
where $\gamma = (1 - v^2)^{-1/2}$ and the observer frame lateral
expansion velocity is then $v_\perp = v'_\perp/\gamma$.  Thus using
eqns.~(\ref{E:Rkdef},\ref{E:Rkeqn},\ref{E:Gammadef}), a given Lorentz
factor, $\Gamma$, and lateral expansion efficiency, $R_k$, one can
determine $\gamma$ and $\gamma'_\perp$.  The total Lorentz factor,
$\Gamma$, is the quantity that is evolved by the dynamical
eqn.~(\ref{E:gamma}).  The angle made by the velocity vector with
respect to the {\it original} radial vector of the shock element, at
$\theta_0$, is
\begin{equation}
\alpha_0 \equiv \arctan \biggl(\frac{v'_\perp}{\gamma v} \biggr) ~.
\label{alphadef}
\end{equation}
In other words, we measure the sideways expansion with
respect to the original radial vector of the shock element; the shock
momentum does not undergo a torque as $\theta$ expands.  The angle
between the shock element velocity and the radial vector at $\theta$
is
\begin{equation}
\alpha = \theta_0 - \theta + \alpha_0
\label{eqnalpha}
\end{equation}
and the shock position, $\theta$, is incremented with radius, $R$, as
\begin{equation}
\theta(R) = \theta_0 + \int \alpha \frac{dR}{R} ~.
\label{eqntheta}
\end{equation}

There are many conceivable elaborations of this model.  For instance,
it is likley that the efficiency of conversion of forward motion to
lateral motion is dependent on Lorentz factor and radius,
$R_k(\gamma,R)$.

This model implies a maximum asymptotic value for the opening angle
$\theta$ which is achieved when the shock has decelerated to
nonrelativistic velocities, at which
eqns.~(\ref{E:Rkdef},\ref{alphadef}) imply $\alpha_0 \approx
\arctan(\sqrt{R_k})$, and since $\alpha$ will become zero at large
radii, then eqn.~(\ref{eqntheta}) gives 
\begin{equation}
\theta_{\text{max}} = \theta_0 + \arctan(\sqrt{R_k}) ~.  
\label{eqn:thmx}
\end{equation}
Thus, as seen in Figs.~\ref{fig:th1gpcmpplot}\&\ref{fig:th0Ngamma},
one does not see exponential growth of the jet opening angle,
i.e.~$\theta \nsim \exp(R)$, and thus the analytically predicted
runaway expansions are physically unrealizable.

\subsection{Amending the Homogeneous Jet Scaling Laws} \label{sec_new_laws}

An interesting result of these calculations is that the afterglow
decay slopes are altered from the theoretical asymptotes both {\em
before} and {\em after} the jet-break time.  In this section I
describe why this occurs and give amended analytical asymptotic
decay slopes.  In Fig.~\ref{fig:th1gpcmpplot} is shown the same
homogeneous, narrow, $\theta_0 = 1^\circ$, jet as in
Figs.~\ref{fig:theta1plot} \& \ref{fig:th1views}, but now with a range
of lateral expansions. A key feature is that, for $\thv = 0^\circ$,
greater lateral expansion is seen to steepen the lightcurve {\em
prior} to the jet break, while not steepening the lightcurve after the
jet break (in fact this slope becomes less steep).  This can also be
seen in the sequence of Figs. \ref{fig:polplot}, \ref{fig:plplot1.6},
\ref{fig:plplot3.7}, \ref{fig:plplot10} for a more `realistic' jet
with $\theta_0 = 5^\circ$.  Thus we calculate behavior {\em opposite}
to that predicted by \citet{rhoads99,sph99}.

To explain this, begin by ignoring Doppler factors, $\delta \propto
\Gamma$, thus from eqn.~(\ref{E:Fnuscale}) the flux varies like
\begin{equation}
F_\nu \propto \Gamma^{4(1+\alpha)} R^3 \theta_A^2
\end{equation}
and the time (eqn.~\ref{E:tobsscale}) 
\begin{equation}
t \propto \frac{R}{\Gamma^2} ~.
\end{equation}
Now parameterizing the Lorentz factor dependence as
\begin{equation}
\Gamma \propto R^{-3\EuScript{A}/2}
\end{equation}
where $\EuScript{A}$ is unity for the case of no lateral expansion and
will be estimated otherwise.  These equations with eqn.~(\ref{E:thetaAdef}) imply
\begin{equation}
F_\nu \propto 
\begin{cases}
  t^{-\frac{1 + 2\alpha - 1/\EuScript{A}}{1 + 1/(3\EuScript{A})}}&   \text{pre-jet-break:}\quad \theta_A \propto 1/\Gamma\\
  t^{-\frac{2(1+\alpha) - 1/\EuScript{A}}{1 + 1/(3\EuScript{A})}}&   \text{post-jet-break:}\quad  \theta_A \propto \theta_0~~.
\end{cases}
\label{E:Fnuscale2}
\end{equation}
In order to estimate the parameter, $\EuScript{A}$, note from
eqn.~(\ref{E:feqn}) that 
\begin{equation}
f(R) \propto \int (\theta R)^2 dR
\label{E:feqn_simple}
\end{equation}
and from eqns.~(\ref{eqntheta},\ref{alphadef}) that 
\begin{equation}
\theta(R) \approx \theta_0 + \int \frac{dR}{\gamma R} ~~.
\label{E:thetaeqn_simple}
\end{equation}
A crucial point here is that the lab-frame angle between the velocity
and radial vectors (eqn.~\ref{alphadef}), $\alpha_0 \propto 1/\gamma$,
and {\em not} $\alpha_0 \propto 1/\Gamma$.  The reason is that maximal
sideways expansion occurs when the shock frame lateral expansion
energy, $\gamma_\perp'$, is roughly the same as the radial expansion
energy, $\gamma$.  From eqn.~(\ref{E:Gammadef}), this implies that
total specific energy of the shock, $\Gamma \sim \gamma^2$.  So if an
expansion angle of $\alpha_0 \sim 1/\Gamma$ is desired, this implies
$\gamma_\perp' \sim \Gamma$ and the total specific energy of the shock
is $\sim \Gamma^2$ which is energetically untenable.  The latter,
erroneous choice of $\alpha_0$ going like $1/\Gamma$ results in the
standard exponential decay of $\Gamma$ with $R$ which was discussed in
the first paragraph of Section \ref{sec_sideways_expansion}.  From
eqn.~(\ref{E:gamma})
\begin{equation}
\Gamma \approx \sqrt{\frac{\Gamma_0}{2 f}}
\label{E:gamma_simple}
\end{equation}
so $1/\gamma \propto 1/\sqrt{\Gamma} \propto f^{1/4}$.  Thus one can
solve eqns.~(\ref{E:feqn_simple},\ref{E:thetaeqn_simple}) for the
asymptotic behaviors.  In the limit of no lateral expansion,
$\theta(R) \rightarrow \theta_0$, and so eqn.~(\ref{E:feqn_simple})
becomes $f \propto R^3$ and thus $\Gamma \propto R^{-3/2}$ so
$\EuScript{A} = 1$ as expected.  For the limiting case when lateral
expansion dominates, eqn.~(\ref{E:thetaeqn_simple}) gives $\theta(R)
\rightarrow \int dR/\gamma/R \propto \int f^{1/4}dR/R$ and
eqn.~(\ref{E:feqn_simple}) becomes $f \propto R^6$ and thus $\Gamma
\propto R^{-3}$ so $\EuScript{A} = 2$.  Therefore, from
eqn.~(\ref{E:Fnuscale2}), for $\alpha= 1/2$ the pre-break light-curve
is steepened from $F_\nu \propto t^{-3/4}$ for negligible lateral
expansion, to $F_\nu \propto t^{-9/7}$ for maximal lateral expansion.
This range nicely brackets the pre-break decays of simulated
lightcurves over a range of lateral expansion rates as seen in the
bottom panel of Figs.~(\ref{fig:polplot}, \ref{fig:plplot1.6},
\ref{fig:plplot3.7}, \ref{fig:plplot10}).  It is also worth noting
that inspection of Figs.~(\ref{fig:plplot10}, \ref{fig:gp10views})
indicates that, not surprisingly, Doppler effects become important for
large lateral expansions.  Such effects were neglected in the
derivation of these scaling laws.

The post-break afterglow decays less steeply than analytical models
predicted.  This is because, as discussed in Section
\ref{sec_sideways_expansion}, a realistic afterglow jet will expand
laterally only in proportion to its forward expansion, thus the
lateral expansion rate decreases at late time (see
Fig.~\ref{fig:th0Ngamma}) and the post-break decay curve will
asymptotically move toward the limit of no lateral expansion;
$\EuScript{A} = 1$.

To summarize, I have implemented a simple model for jet dynamics in
which the lateral expansion of the jet is an effect of forward
expansion.  I argue that this basic cause/effect relationship is a
necessary component of any model for jet dynamics.  Resulting jet
simulations show that lateral expansion {\it i)} smooths the break in
the jet lightcurve, {\it ii)} steepens the pre-break light curve, but
does {\it not} significantly steepen the post-break slope of the
light-curve, {\it iii)} can significantly alter the shape of a
lightcurve.  Thus I suggest that the oft-cited scaling laws described
at the beginning of this section do not accurately represent the
evolution of a laterally expanding jet.  To properly diagnose
lightcurve slopes and breaks, simulations such as those shown here are
necessary.  A preliminary conclusion that can be drawn from the
simulations shown in this section is that large efficiencies in
conversion of forward kinetic energy to lateral kinetic energy,
i.e.~$R_k \gtrsim 0.1$, are probably not physical, in the context of
the homogeneous jet model, because the lightcurves they produce bear
little resemlance to observed lightcurves.  In this way one can begin to
constrain the physics of afterglow shocks with observations.

\section{The Structured, Universal Jet}

The primary purpose of the numerical framework discussed thus far is
to quantitatively study the afterglow lightcurves from structured
jets, i.e.~with a non-uniform energy or velocity distribution.  As
suggested by \citet{rlr02} and \citet{zm02}, an asymptotic decrease in
energy per solid angle like $\EuScript{E}(\theta) \propto \theta^{-2}$
can reproduce the observed relation $E \propto t_j^{-1}$
\citep{fksd01}.  Furthermore, allowing $\EuScript{E} \sim$ constant
for angles within a core, $\theta < \theta_c$ provides a natural
explanation for the dearth of small angle jets, less than $3^\circ$
also reported by \citet{fksd01}.  As such I define an ansatz energy
profile
\begin{equation}
\EuScript{E}(\theta) = \frac{\EuScript{E}_{52}}{1 + (\theta/\theta_c)^2} ~\frac{10^{52}}{4\pi} ~ \text{ergs sr}^{-1}~.
\label{energy_profile}
\end{equation}

How does one choose an initial $\Gamma_0(\theta)$?  based on
simulations by \citet{zwm03}, $\Gamma_0 \sim $ const.~and so this
choice will be used for the structured jet runs in this paper.
However, it is important to realize, and the following simulations
confirm this, that initial $\Gamma_0$ only determines the deceleration
time of the shock, but subsequent evolution is independent of it. To
see this, noting the global energy has dependence $\EuScript{E}
\propto \Gamma^2 R^3$ from eqns.~(\ref{E:enmomeqns}), then for any
radius greater than the deceleration radius, $R > R_d$, the Lorentz
factor is determined by $\Gamma \propto \EuScript{E}^{1/2}$, independentof initial $\Gamma_0$.

\begin{figure}
\epsscale{1.}
\plotone{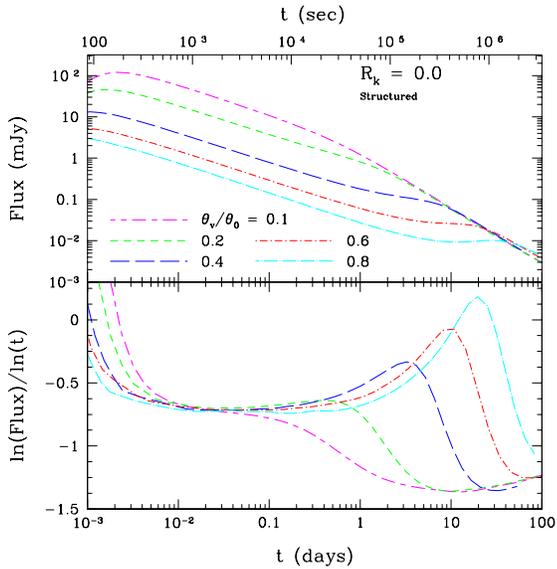} 
\caption{Lightcurves for a structured jet with isotropic equivalent
energy $10^{53}$ ergs, $\EuScript{E}_{52} = 10$, using
eqn.~(\ref{energy_profile}), and initial Lorentz factor $\Gamma_0 =
100$ everywhere.  One sees corroboration of the basic thesis of
\citet{rlr02}, with some flattening at larger viewing angles. }
\label{fig:gp1th30plot}
\end{figure}

In Fig.~\ref{fig:gp1th30plot} is shown a series of lightcurves from a
universal structured jet with isotropic equivalent energy
$\EuScript{E} = 10^{53}$ ergs$/4\pi$, initial Lorentz factor
$\Gamma_0(\theta) = 100$, core angle $\theta_c = 3^\circ$ and
$\theta_0 = 30^\circ$.  These parameters generally reflect those
inferred from burst observations by \citet{pk02}.  This figure
demonstrates the characteristics of a lightcurve derived from the
structured jet described by \citet{rlr02}.

Using the scaling of eqns.~(\ref{E:Fnuscale}, \ref{E:tobsscale}) with
$\EuScript{E} \propto \gamma^2 R^3$ one finds
\begin{equation}
F_\nu \propto \Gamma^{3\alpha-1} \delta^{\alpha+3} \theta_A^2 \EuScript{E}
\label{E:structFscale}
\end{equation}
and
\begin{equation}
t \propto \EuScript{E}^{1/3} \Gamma^{-5/3} \delta^{-1} 
\label{E:structtscale}
\end{equation}
where the asymptotic limit of $\theta_A$ is given in Table
\ref{t:simplejet1} and $\delta \approx 2\Gamma/(1+\theta^2 \Gamma^2)
\sim \Gamma$ for $\theta \lesssim 1/\Gamma$.  Using
eqn.~(\ref{E:structtscale}) and $\EuScript{E} \propto
(\theta_c/\thv)^{-2}$ for a structured jet, at an early time
($\Gamma \gg 1/\thv$) the Lorentz factor scales with observed angle as
$\Gamma \propto \thv^{-1/4}$ and so $F_\nu \propto
\Gamma^{4\alpha}\EuScript{E} \propto \thv^{-(\alpha+2)}$ so the flux
at a given time depends on viewing angle like
\begin{equation}
F_{\nu,\text{simultaneous}} \propto \thv^{-(\alpha + q)} \propto \thv^{-5/2}
\end{equation}
for $\alpha = 1/2$, and $q = 2$.  

There can be seen a gradual flattening, and eventual appearance of a
bump, in the lightcurves of Fig.~\ref{fig:gp1th30plot} with increasing
viewing angle $\thv$.  This can be understood when one considers that
the early phase of the lightcurve, i.e.~$\Gamma > 1/\thv$, is
dominated by emission primarily along the line of sight (los) with energy
$\EuScript{E} \propto (\theta_c/\thv)^{-2}$, while the flux at the
break is dominated by the energetic core moving at angle $\thv$ with
respect to the observer.  Here I show that these two components have
distinct laws for their breaks, and this difference between these two
laws makes each component distinct for large viewing angles, $\thv$,
thus creating a bump.

The line of sight component can be modeled as a homogeneous jet with
opening angle $\thv$.  As such one expects a break when $\Gamma \sim
\delta \sim 1/\theta_A \sim 1/\thv$, so $F_{\nu,\text{los}} \propto
\Gamma^{4\alpha} \EuScript{E} \propto \thv^{-(4\alpha+q)}$.
Furthermore $t_{\text{los}} \propto \EuScript{E}^{1/3} \Gamma^{-8/3}
\propto \thv^2$.  So
\begin{equation}
F_{\nu,{\text{los}}} \propto t_{\text{los}}^{-1/2(4\alpha+q)} \propto t^{-2}_{\text{los}}
\label{E:Flos}
\end{equation}
where $\alpha=1/2$ and $q=2$.  The jet core component can be modeled as
a narrow jet seen far off-axis; $\thv \gg \theta_A$ where $\theta_A =
\theta_c$.  One expects a maximum flux when $\Gamma \sim 1/\thv$, thus
$F_{\nu,\text{core}} \propto \Gamma^{4\alpha+q} \EuScript{E} \propto
\thv^{-(4\alpha + q)}$, where $\EuScript{E} = 1$ for the core.  It
follows that $t_{\text{core}} \propto \EuScript{E}^{1/3} \Gamma^{-8/3}
\propto \thv^{8/3}$.  Thus the core is brightest at
\begin{equation}
F_{\nu,{\text{core}}} \propto t^{-3/8(4\alpha+q)}_{\text{core}} \propto t^{-3/2}_{\text{core}}
\label{E:Fcore}
\end{equation}
for $\alpha=1/2$ and $q=2$.  This relation traces the jet break in the
lightcurve, $F_\nu(t)$, over a range of viewing angles.  The different
decay laws for eqns.~(\ref{E:Flos}, \ref{E:Fcore}) demonstrate why
both components of a lightcurve from a structured jet will diverge for
large break times.  Seen another way:
\begin{equation}
\frac{t_{\text{core}}}{t_{\text{los}}} \propto \thv^{2/3} ~,
\end{equation}
thus the contribution to the lightcurve from the core happens progressively later than that of the line-of-sight material.  

\subsubsection{The ``Bump''}

\begin{figure}
\plotone{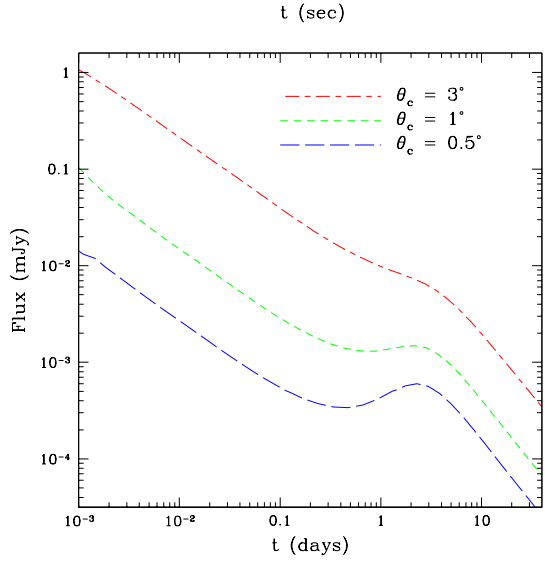} 
\caption{The bump in the lightcurves at the jet break time becomes
more prominent with smaller cores. These afterglows have $E_{iso} =
10^{52}$ ergs, $\gamma_0 = 100$, $v_\perp = 0$, and are seen at $\thv
= 12^\circ$. }
\label{fig:bump}
\end{figure}

As discussed above, under certain conditions a ``bump'' can appear in
the lightcurve at the jet-break time.  This typically happens at large
viewing angles and small jet cores.  For example, Fig.~\ref{fig:bump}
shows three lightcurves from identical afterglows except for variation
in the size of the jet core, $\theta_c$, demonstrating that a narrower
core produces a more prominent bump.  This bump can be explained by
considering how the flux varies as the observer sees an
ever-increasing ($\sim 1/\gamma$) angular region of a surface with
varying energy density.  It can be shown that the observed flux varies
like $F \propto \gamma^p \EuScript{E}$ where $p=2$ for $1/\gamma <
\theta_0$ (isotropic) and $p=4$ otherwise (jet-like).  Also, the
energy per steradian of the afterglow goes like $\epsilon \propto
(\theta_c/\theta)^{q}$ where in this case $q=2$.  Thus the flux will
vary like
\begin{equation}
F(1/\gamma) \propto \biggl(\frac{1}{\gamma}\biggr)^{-p} \biggl(\frac{\theta_c}{\theta}\biggr) \biggl(\frac{\theta_c}{\theta - 1/\gamma}\biggr)^{q-1}
\label{E:bumpflux}
\end{equation}
so that early in the evolution, when $1/\gamma \ll \theta$, the flux
is down by $(\theta_c/\theta)^q$ as expected, but when the beaming
angle has expanded to reach the core, i.e.\ $1/\gamma = \theta -
\theta_c$, the flux averaged over the surface will be $\propto
(\theta_c/\theta)^{q-1}$.  The flux of eqn.~(\ref{E:bumpflux})
decreases as $\gamma$ decreases until $1/\gamma = p/(p+q-1) \theta$.
So the smaller $q/p$ is, the farther out the bump will occur.  The
condition for {\em no} bump to be observed is when the jet core
becomes visible, i.e.\ $1/\gamma = \theta - \theta_c$, the flux has
not yet begun to increase, thus
\begin{equation}
\theta_c > \frac{q-1}{p+q-1} \theta ~.
\end{equation}
So for $p=2$ and $q=2$ we have $\theta_c > \theta/3$ and so a bump
will be visible if the jet is viewed at angles, $\theta$, much in
excess of $3 \theta_c$.  This analysis is born out in the
simulations and gives constraints on the range of allowed viewing
angles, $\theta$, with respect to the size of jet core, $\theta_c$,
and the steepness of the jet decay structure, $q$.

There are several examples of bumps in light- curves including GRB
970228, GRB 970508, GRB 980326 \cite[for a discussion, see][]{zm02b}.
While in general gamma-ray burst afterglows do not exhibit bumps in
their lightcurves just prior to the break, GRB 000301C exhibited a
prominent, achromatic bump at the jet-break time which has been
interpreted as a gravitational lensing event \citep{gls00} and
alternatively as continuous energy injection by a millisecond pulsar
\citep{zm01}.  In light of the present calculations, it is possible
that the bump in the lightcurve of GRB 000301C was due to a simple
perspective effect onto a jet with a narrow core or a steep decay
curve.  This explanation is appealing in that it does not require
external mediators (i.e. a lensing body or a pulsar) to create it.  If
the bump in GRB 000301C and possibly those of other burst lightcurves
could be positively attributed to perspective onto a structured
afterglow, key information could be determined about the size
(i.e.~the jet core $\theta_c$) and shape (the decay structure $q$) of
the afterglow.

\subsection{Uniform Lateral Expansion} \label{section:uniform}

Having constructed a structured jet model, we are interested in the
effects of lateral expansion on said model.  Since the evolution of
the afterglow shock is uncertain, I choose to study two basic
paradigms.  First, in this section I simply apply uniform expansion to
the afterglow as was done for the homogeneous jet in the previous
section. Thus expansion is governed by eqn.~(\ref{E:Rkeqn}).  The second
model, discussed in the next section, employs nonuniform expansion,
where the hotter core expands faster than the cooler wings of the
afterglow. 

\begin{figure}
\plotone{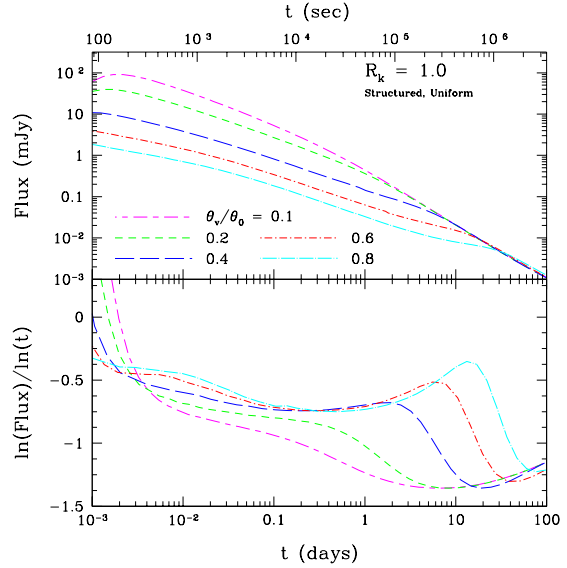} 
\caption{Lightcurves for the structured jet of
Fig.~\ref{fig:gp1th30plot} with large, uniform lateral expansion, $R_k
= 1$.  Lateral expansion suppresses and smooths the bump in the
lightcurve at the jet break time for large viewing angles, $\thv$.}
\label{fig:gp10th30plot}
\end{figure}

Examination of the progression of increasing lateral expansions
Figs.~\ref{fig:gp1th30plot}, \ref{fig:gp3.7th30plot} \&
\ref{fig:gp10th30plot} shows that lateral expansion suppresses the
bump at the jet break time viewed at large $\thv$.  However, this also
makes the breaks at small $\thv$ less pronounced.  This behavior,
where the jet break is more pronounced at high viewing angles, $\thv$,
than at small ones, is quite general.  Even a nonuniform expansion
paradigm discussed in the next section demonstrates this.

\subsection{Nonuniform Lateral Expansion} \label{section:nonuniform}

\begin{figure}
\plotone{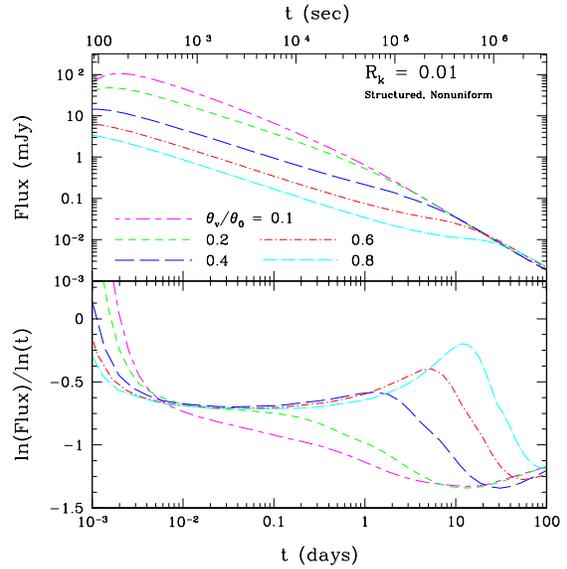} 
\caption{Lightcurves for a structured jet with a nonuniform lateral
expansion (eqn.~\ref{E:bernoulli}).  Other parameters are the same as
in Fig.~\ref{fig:gp1th30plot}.  One can see that even a rather small,
$R_k = 0.01$, nonuniform lateral expansion can affect the lightcurves
when compared to $R_k=0$ in Fig.~\ref{fig:gp1th30plot}. }
\label{fig:gp1.6th30epsplot}
\end{figure}

Uniform expansion of the structured jet is likely oversimplified.  A
more accurate sideways expansion perscription should encapsulate one's
hydrodynamic intuition that a fluid element will be accelerated
proportionally to the gradient of its internal energy density. A simple model with this characteristic is
\begin{equation}
R_k(\theta) = R_{k0} \biggl(1 - \frac{\EuScript{E}(\theta)}{\EuScript{E}_{52}} \biggr) = R_{k0} \frac{(\theta/\theta_c)^2}{1 + (\theta/\theta_c)^2}
\label{E:bernoulli}
\end{equation}
using eqn.~(\ref{energy_profile}) and where $\theta_c$ is not
constant, but allowed to laterally expand with the jet by staying
assigned to a particular fluid element.  Notice that
eqn.~(\ref{E:bernoulli}) replicates the uniform expansion of
eqn.~(\ref{E:Rkeqn}) for small angles, $\theta \ll \theta_c$, but
transitions to a constant, rigid-body, expansion for large angles.
Figs.~\ref{fig:gp1.6th30epsplot} \& \ref{fig:gp3.7th30epsplot}
demonstrate that non-uniform expansion can be very effective at
creating a very sharp break in the lightcurve.

A key approximation of eqn.~(\ref{E:bernoulli}) is that it assumes a
fixed functional form for $R_k$ even as the shock surface evolves.
This represents a model in which the proper hydrodynamic timescale,
which increases with time, become longer than the deceleration
timescale, thus the early time expansion morphology is ``frozen in''
and predominantly determines the subsequent evolution.  Further work
is required to determine the hydrodynamical evolution of the
afterglow.

\section{Discussion}

I have implemented a simple model for afterglow jet spreading in which
lateral expansion depends upon energy from forward expansion.  This
model demonstrates that a homogeneous jet does not exhibit a lateral
expansion dominated phase, as described by analytical arguments
\citep{rhoads99,sph99}.  In particular, dynamically $\gamma \nsim
\exp(-R)$, (Fig.~\ref{fig:th0Ngamma}) and observationally $F_\nu \nsim
t^{-p}$ after the break, (Figs.~\ref{fig:plplot1.6},
\ref{fig:plplot3.7}, \ref{fig:plplot10}).  The only source of a break
in this model is the observation of the physical edge of the jet.  As
shown in Figs.~\ref{fig:plplot1.6}, \ref{fig:plplot3.7},
\ref{fig:plplot10}, this break is `sharpened' by lateral expansion in
that it becomes less dependent on viewing angle.

\begin{figure}
\plotone{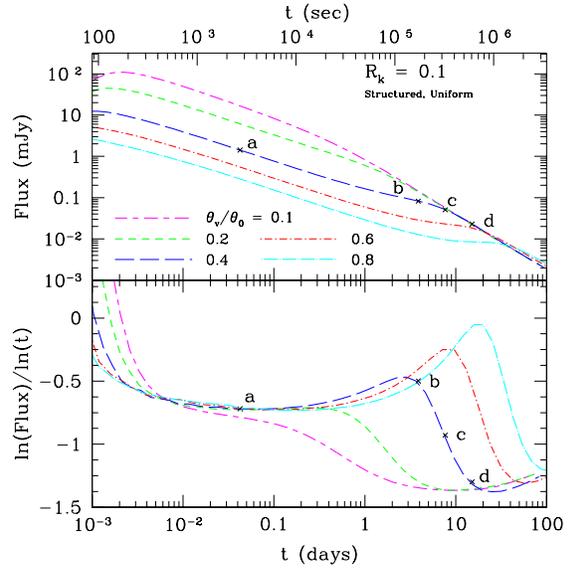} 
\caption{Lightcurves for the same structured jet as
Fig.~\ref{fig:gp1th30plot}, but with moderate uniform lateral
expansion, $R_k = 0.1$. Notice a slight suppression of the bump at the
jet-break time for large $\thv/\theta_0$ compared to
Fig.~\ref{fig:gp1th30plot}.  Fig.~\ref{fig:fig100d12gp3.7th30c3c}
shows selected flux contours for $\thv = 12^\circ$.}
\label{fig:gp3.7th30plot}
\end{figure}

\begin{figure}
\plotone{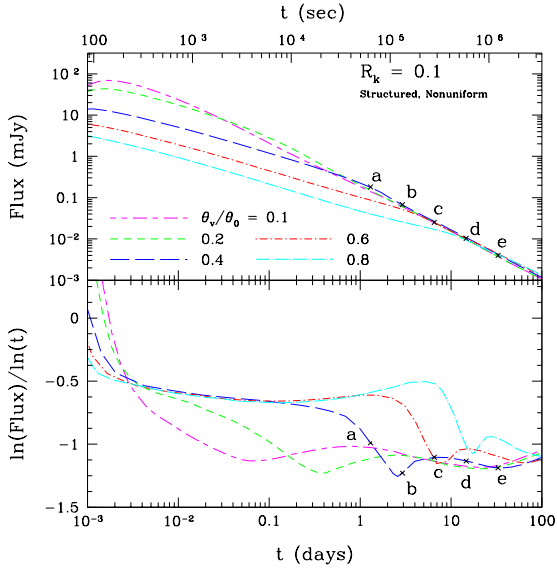} 
\caption{ The modest nonuniform expansion shown here, $R_k = 0.1$,
compared to Figs.~\ref{fig:gp1.6th30epsplot} \& \ref{fig:gp1th30plot}
makes a very sharp break with no bump for $\thv > 0.2 \theta_0$, but
washes out the break as seen at smaller viewing
angles. Fig.~\ref{fig:figepsplot} shows selected flux contours for
$\thv = 12^\circ$. }
\label{fig:gp3.7th30epsplot}
\end{figure}

Also, I have studied structured jets as seen at various viewing angles
and with various sideways expansions.  This study confirms the key
results of \citet{rlr02,zm02}, $t_j \propto \theta_v^{8/3}$ and
$E_{iso} \propto t_j^{-1}$.  Furthermore, the jet break is seen to be
sharp and can exhitibit some flattening and even a bump prior to the
break at large viewing angles compared to the core size, $\thv \gg
\theta_c$.  It appears to be a general feature of the universal jet
model that late time jet breaks (i.e.~large $\thv$) are more
pronounced than early time breaks.  That is, if early time breaks are
sharp, then late time breaks will tend to have a flattening or a bump
(e.g.~Fig.~\ref{fig:gp3.7th30plot}), but if this bump is quenched,
perhaps by nonuniform spreading of the jet, the early-time breaks
become weak or are washed out altogether
(e.g.~Fig.~\ref{fig:gp3.7th30epsplot}).  In general, the sharpness of
the structured jet break is determined by the size of the core, $t
\propto \theta_c^{8/3}$, and is thus intrinsically sharper than the
break in the homogeneous jet model which depends upon the outside edge
$t \propto \theta_0^{8/3}$.  As such the structured jet can explain
sharper breaks than the homogeneous jet model.

Future work with E.~Rossi et al.~will focus on polarization
(e.g.~Fig.~\ref{fig:polplot}) as a tool to discriminate between the
homogeneous and structured jet paradigms.  Also, improved understanding
of the evolution of the jet, whether it be hydrodynamic or otherwise,
will allow for more quantitative expansion models and more predictive
simulations.

\acknowledgments

I wish to convey my sincerest thanks to E. Ramirez-Ruiz, E. Rossi,
D. Lazzatti and M. Rees for their kind hospitality at the U. of
Cambridge and insightful, enjoyable discussions during the preparation
of this document.  I also wish to thank T. Galama and R. Sari for
stimulating, useful discussions regarding this work.  This work was
performed under the auspices of the U.S. Department of Energy by
University of California Lawrence Livermore National Laboratory under
contract W-7405-ENG-48.


\begin{figure}
\epsscale{1.} 
\plotone{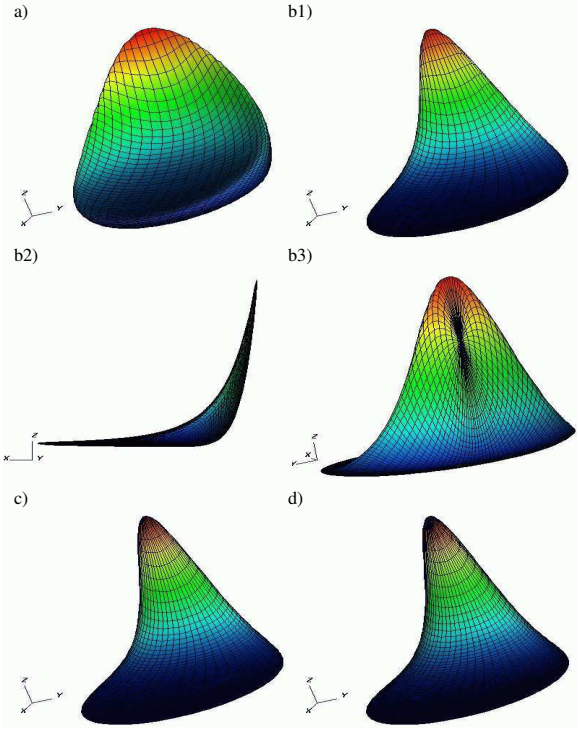} 
\caption{ Flux contours for selected points of
Fig.~\ref{fig:gp3.7th30plot} for $\thv = 12^\circ$.  At early times,
a), the flux surface is distorted from a symmetric bowl (see
Fig.~\ref{fig:theta1plot} b) by both lateral expansion, as in
Fig.~\ref{fig:gp10views} a), and the intrinsic structure of the jet.
Near the jet-break time plots b) show different views of the flux
surface.  Plot b3) shows the physical pole of the afterglow nearly
coincident with the flux peak. This demonstrates the origin of the
break in the lightcurve in the structured jet model: the coincidence
of the peak flux rim and the actual pole of the jet.  As such the edge
of the jet does not play a role in the jet break as it does in the
homogeneous jets. Plots c) and d) show this coincidence clearly.  Grid
resolutions are $\Delta \theta = 30^\circ/300$, $\Delta \phi =
360^\circ/360$ for a) and $\Delta \theta = 30^\circ/100$, $\Delta \phi
= 360^\circ/60$ for b), c) and d).}
\label{fig:fig100d12gp3.7th30c3c}
\end{figure}

\begin{figure}
\epsscale{1.} 
\plotone{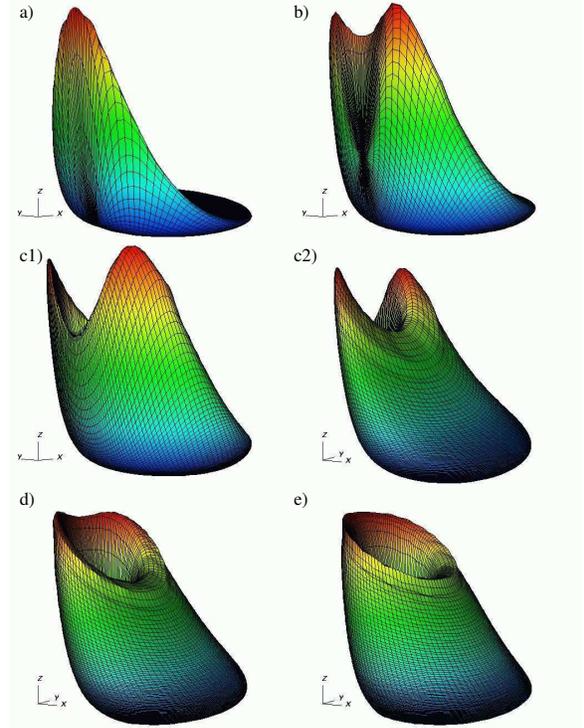} 
\caption{Selected flux surfaces for a structured jet with a nonuniform
expansion paradigm at $\thv = 12^\circ$. Grid resolutions are $\Delta
\theta = 30^\circ/100$ and $\Delta \phi = 360^\circ/120$. Prior to the
break, a), the morphology is similar to the uniform expansion
(Fig.~\ref{fig:fig100d12gp3.7th30c3c}). The hot core of the jet expans
and decelerates more rapidly than the wings. Thus a crater appears as
the core comes into view (b - e). This sharp flux deficit makes a
sharp jet-break (Fig.~\ref{fig:gp3.7th30epsplot}). }
\label{fig:figepsplot}
\end{figure}



\end{document}